\documentclass[12pt,epsf]{article}
\usepackage{graphicx}
\usepackage{amsmath}
\usepackage{amssymb}

\begin{document}
\def\a{\alpha}
\def\b{\beta}
\def\c{\varepsilon}
\def\d{\delta}
\def\e{\epsilon}
\def\f{\phi}
\def\g{\gamma}
\def\h{\theta}
\def\k{\kappa}
\def\l{\lambda}
\def\m{\mu}
\def\n{\nu}
\def\p{\psi}
\def\q{\partial}
\def\r{\rho}
\def\s{\sigma}
\def\t{\tau}
\def\u{\upsilon}
\def\v{\varphi}
\def\w{\omega}
\def\x{\xi}
\def\y{\eta}
\def\z{\zeta}
\def\D{{\mit \Delta}}
\def\G{\Gamma}
\def\H{\Theta}
\def\L{\Lambda}
\def\F{\Phi}
\def\P{\Psi}

\def\S{\Sigma}

\def\o{\over}
\def\beq{\begin{eqnarray}}
\def\eeq{\end{eqnarray}}
\newcommand{\gsim}{ \mathop{}_{\textstyle \sim}^{\textstyle >} }
\newcommand{\lsim}{ \mathop{}_{\textstyle \sim}^{\textstyle <} }
\newcommand{\vev}[1]{ \left\langle {#1} \right\rangle }
\newcommand{\bra}[1]{ \langle {#1} | }
\newcommand{\ket}[1]{ | {#1} \rangle }
\newcommand{\EV}{ {\rm eV} }
\newcommand{\KEV}{ {\rm keV} }
\newcommand{\MEV}{ {\rm MeV} }
\newcommand{\GEV}{ {\rm GeV} }
\newcommand{\TEV}{ {\rm TeV} }
\def\slash#1{\ooalign{\hfil/\hfil\crcr$#1$}}
\def\diag{\mathop{\rm diag}\nolimits}
\def\Spin{\mathop{\rm Spin}}
\def\SO{\mathop{\rm SO}}
\def\O{\mathop{\rm O}}
\def\SU{\mathop{\rm SU}}
\def\U{\mathop{\rm U}}
\def\Sp{\mathop{\rm Sp}}
\def\SL{\mathop{\rm SL}}
\def\tr{\mathop{\rm tr}}

\baselineskip 0.7cm

\begin{titlepage}

\begin{flushright}
UCB-PTH-07/19

UT-07-28

IPMU 07-0001
\end{flushright}

\vskip 1.35cm
\begin{center}
{\large \bf Moduli Stabilization in Stringy ISS Models}
\vskip 1.2cm

Yu Nakayama${}^{1}$, Masahito Yamazaki${}^{2}$ and T.T. Yanagida${}^{2,3}$

\vskip 0.8cm

${}^1${\it Berkeley Center for Theoretical Physics and Department of Physics, 
\\
University of California, Berkeley, California 94720-7300}

\vskip 0.4 cm

${}^2$
{\it  Department of Physics, University of Tokyo,\\
     Tokyo 113-0033, Japan}

\vskip 0.4 cm

${}^3$
{\it Institute for the Physics and Mathematics of the Universe, University of Tokyo, \\
Kashiwa 277-8568, Japan}

\vskip 3.5cm

\abstract{We present a stringy realization of the ISS metastable SUSY breaking model with moduli stabilization. The mass moduli of the ISS model is stabilized by gauging of a $U(1)$ symmetry and its D-term potential. The SUSY is broken both by F-terms and D-terms. It is possible to obtain  de-Sitter vacua with a vanishingly small cosmological constant by an appropriate fine-tuning of flux parameters.}
\end{center}
\end{titlepage} 

\setcounter{page}{2}

\section{Introduction}
One of the central dogmas in phenomenological applications of supersymmetry (SUSY) is its dynamical breaking because it will provide a natural solution to the hierarchy problem. Quite a few models within the field theory have been proposed and investigated from various viewpoints (see e.g. \cite{Shadmi:1999jy}\cite{Intriligator:2007cp} for reviews). Given their success, string theoretic realization of such dynamical SUSY breaking models is of great significance but the attempts have been successful almost exclusively in a non-compact global SUSY limit, where the gravity decouples.\footnote{With the use of flux compactification and nonperturbative effects, it is possible to obtain string compactification with broken SUSY based on the approach initiated in \cite{Kachru:2003aw}. However, the connection to the field-theory dynamical SUSY breaking models is far from obvious.} 

In the global SUSY limit, metastable SUSY breaking vacua are now known ubiquitous in string theory \cite{OO}. However, almost all such realizations from the string theory have been done in the global SUSY limit so far, where the gravity decouples and the moduli are fixed by hand. One simple example of such metastable SUSY breaking models is the SQCD with small mass deformation recently proposed by Intriligator, Seiberg and Shih (ISS) \cite{Intriligator:2006dd}, which breaks the SUSY by the F-term vacuum expectation value. It is easy to embed the ISS model and its variants in the string theory, but only in the global SUSY limit.\footnote{See e.g. \cite{ISSstring} for string construction in the global SUSY limit. Some attempts to stabilize K\"ahler moduli in the string compactification with the ISS model can be found in \cite{Dudas:2006gr,Abe:2006xp,Lebedev:2006qc,Lalak:2007qd,GomezReino:2007qi,Abe:2007yb,Brax:2007xq}, but with fixed ISS mass moduli. In the heterotic string compactification, the stabilization of the mass moduli was discussed in \cite{Serone:2007sv}.} Once we couple them to the supergravity, few viable models are known.

Indeed, it is not clear whether or not the ISS(-like) models are in the swampland \cite{Vafa:2005ui}\cite{Ooguri:2006in} once we introduce finite gravitational coupling. More generally, any SUSY breaking models in the global SUSY limit (e.g. \cite{Dymarsky:2005xt}\cite{Dine:2006gm}\cite{A1}\cite{Aharony:2007db}\cite{A2}\cite{A3} in addition to the ISS model) suffer the same problem of moduli stabilization because typically the order parameter of the SUSY breaking tends to be zero once such a parameter becomes dynamical.

In this paper, we give a concrete solution to the moduli stabilization problem in the string compactification with the dynamical SUSY breaking. We use an explicit stringy realization of the ISS(-like) model because of its simplicity. Our key idea to stabilize the moduli is to introduce a dynamical Fayet-Iliopoulos (FI) term. The dynamical FI term itself will also be stabilized by the competition between the F-term potential and the D-term potential.

In our construction, the SUSY is broken both by F-terms and D-terms, and we can obtain a de-Sitter vacuum by an appropriate fine-tuning of flux parameters. The necessity of the D-term has an interesting phenomenological consequence: one may be able to realize the strongly coupled D-term gauge mediation \cite{Nakayama:2007cf},\footnote{For an F-term gauge mediation, see \cite{MN}.} where very light gravitino ($\sim O(1)$ eV) can be realized with a possible natural candidate for the dark matter \cite{Hamaguchi:2007rb}.

The organization of the paper is as follows. In section 2, we present our scheme of moduli stabilization of the mass moduli and the dynamical FI term in our ISS(-like) model within the global SUSY limit. In section 3, we study a stringy realization of the ISS(-like) model with the moduli stabilization. In section 4, we present some discussions for future studies.

\section{Moduli stabilization of ISS model in global SUSY limit}
We would like to begin with the moduli stabilization problem of the ISS model \cite{Intriligator:2006dd} in the global SUSY limit. The original ISS model consists of $SU(N_c)$ SQCD with $N_f$ pairs of fundamental quarks $\varphi_i$ and $\bar{\varphi}^i$, where $i$ runs from 1 to $N_f$ with $ N_c<N_f < \frac{3}{2} N_c$ so that the Seiberg dual of the SQCD is in the infrared-free regime. We introduce a small mass-superpotential perturbation
\begin{eqnarray}
W_{\mathrm{electric}} = m \varphi_i\bar{\varphi}^i \ ,
\end{eqnarray}
where the mass parameter $m$ should be much smaller than the dynamical scale $\Lambda$: $|m| \ll |\Lambda|$ to achieve a calculable metastable vacuum.

The metastable vacuum appears in its clearest form in the Seiberg dual description \cite{Seiberg:1994pq}. We have $SU(N_f-N_c)$ gauge theory with $N_f$ pairs of dual fundamental quarks $q_i$ and $\bar{q}^i$ and singlet meson superfields $M_{ij} = \varphi_i \bar{\varphi}_j$. The dual theory has a superpotential
\begin{eqnarray}
W_{\mathrm{magnetic}} = m \mathrm{Tr}M + \frac{1}{\mu} q^iM_{ij} \bar{q}^j + \mathrm{nonperturbative  \ term} \ .
\end{eqnarray}
We will neglect the nonperturbative term hereafter by focusing on the metastable SUSY vacuum.
From the rank condition, the model breaks SUSY with the potential
\begin{eqnarray}
V = N_c |m|^2 |{\Lambda}|^2 \label{potISS}
\end{eqnarray}
up to  a numerical constant of order 1 by setting $M=0$, $q= \bar{q} = i \sqrt{m\mu} \mathbf{1}_{N_f-N_c\times N_f- N_c}$. It can be further shown that the vacuum is metastable with a sufficiently long life-time as long as $|m| \ll |\Lambda| $ \cite{Intriligator:2006dd}.

Now, the problem of any spontaneous SUSY breaking is moduli stabilization as discussed in the introduction. In the string construction that we will investigate in the next section, the mass parameter $m$ becomes dynamical, which will be denoted by $\rho$. As is clear from the potential $\eqref{potISS}$, SUSY is restored by setting $\rho = 0$. 

Our idea to stabilize the moduli $\rho$ is to introduce an extra anomalous $U(1)$ gauge symmetry with a possibly field dependent FI parameter $\xi$. 
Under the $U(1)$ symmetry, we assign charge $-2$ to $\rho$ and $+1$ to both $\varphi$ and $\bar{\varphi}$. The quantum anomaly demands that $\Lambda^{3N_c-N_f}$ should be charged with charge $2N_f$. On the dual side, we assign charge $+2$ to $M$. The charge of the dual Landau-pole scale $\tilde{\Lambda}$ and the charge of the dual quarks are determined from the following constraint: we have a schematic relation between the baryon superfields \cite{Intriligator:1995au}
\begin{eqnarray}
B\sim \varphi^{N_c} = \left(-(-\mu)^{N_c-N_f} \Lambda^{3N_c-N_f}\right)^{1/2} q^{N_f-N_c} 
\end{eqnarray}
and we also demand the $U(1)$ invariance of the dual superpotential $W_{\mathrm{magnetic}} = \frac{1}{\mu} q_iM^{ij}\bar{q}_j$.\footnote{We have a further relation $\Lambda^{3N_c-N_f} \tilde{\Lambda}^{2N_f-3N_c} = (-1)^{N_f-N_c}\mu^{N_f}$, but it does not give an extra condition.}
With a further natural assumption that $\mu$ is neutral under the $U(1)$,\footnote{This is also natural from the string construction, where the local Yukawa interaction does not depend on the volume of the 4-cycle, at least in non-compact examples. In any case, we can always absorb the charge of $\mu$ by field redefinition of $q$ and $\bar{q}$.}  we see that the charge of the dual quark $q$ and $\bar{q}$ is $-1$ and $\tilde{\Lambda}^{2N_f-3N_c}$ has  $-2N_f$ of charge. As we will discuss later, the anomaly will be cancelled by four-dimensional Green-Schwartz mechanism, which is also a natural consequence of the string construction of the model. 

With this charge assignment, we have an extra $D$-term contribution to the potential
\begin{eqnarray}
V_D = \frac{g^2}{2} \left(\xi - |q|^2 - |\bar{q}|^2 -2|\rho|^2 + 2\frac{|M|^2}{|\tilde{\Lambda}|^2}\right)^2  + (\mathrm{higher\ Kahler\ corrections}) \ .
\end{eqnarray}
As is the case with the original ISS model, $q$, $\bar{q}$ and $M$ are fixed (up to remaining symmetries) for a given $\rho$ with small $g$. This can be also done by solving the equation of motions for the matter fields. An additional $F$-term potential $|\partial_\rho W|^2$ makes $M$ vanish, which is consistent with the ISS vacuum. The total effective potential for the $\rho$ field is, then, given by
\begin{eqnarray}
V= V_F+V_D = N_c|\rho|^2 |\Lambda|^2 + \frac{g^2}{2} \left(\xi - 2|\rho|^2 -2(N_f-N_c)|\mu \rho|\right)^2 \ . \label{topote}
\end{eqnarray}
For small $\xi$, which is necessary to obtain $|\rho| \ll |\Lambda|$, the total potential is minimized at $|\rho| = \frac{g^2|\mu|\xi(N_f-N_c)}{N_c|\Lambda|^2 + 2g^2|\mu|^2(N_f-N_c)^2}$. In this way, the mass moduli $\rho$ can be stabilized at a value consistent with the ISS metastable vacuum, and $|\rho|$ acquires a mass of order $|\Lambda|$ (or $g|\mu|$). Due to the vacuum expectation value of $|\rho|$, all the matter fields acquired the mass squared of order $\frac{|\Lambda|^2|\rho|}{|\mu|} \sim g^2 \xi$.\footnote{Our approximate way to stabilize the moduli is based on $g^2$ expansion. Since the matter fields $q,\bar{q}$ are lighter than the mass moduli $\rho$, our strategy to minimize ISS matter fields first might need corrections. We have, however, checked the analytic, as well as numerical, solution of the equation of motions to see that the deviation is small as long as $g^2$ is small.}

This is not the end of the story, however. As we have noticed, the gauging of this $U(1)$ is anomalous,\footnote{Non-anomalous gauging would be obtained by gauging the baryon symmetry of the ISS model. However, under the baryon symmetry, $m$ (or $\rho$) is not charged, and hence does not lead to the moduli stabilization we aim at.} and we have to implement the four-dimensional Green-Schwartz mechanism  \cite{Dine:1987xk}. Furthermore, when the FI parameter becomes dynamical, the SUSY vacuum is restored by relaxing the D-term. 

These two problems are solved at once when we consider the string model. From the field-theory viewpoint in the global SUSY limit, things go as follows. The coupling constant of the model becomes a chiral superfield $T(x;\theta) = \frac{1}{g^2}(x) + \frac{i}{8\pi^2}\phi(x) + O(\theta) $ and the axion part transforms as $\phi(x) \to \phi(x) -2N_f\alpha(x)$ under the gauge transformation $A_\mu(x) \to A_\mu(x) + \partial_\mu \alpha(x)$ to cancel the anomaly. The K\"ahler potential, therefore, should depend on the gauge invariant combination $T+T^\dagger -\frac{N_f}{4\pi^2} V$, where $V$ is the vector superfield corresponding to the $U(1)$ gauge group.  The action contains both the dynamical FI-term and the Higgs-term:
\begin{eqnarray}
\int d^4\theta K(T+T^\dagger -\frac{N_f}{4\pi^2}V) = \left(\frac{\partial K}{\partial V}\right)_{V=0} V|_{\theta^4} + \frac{1}{2}\left(\frac{\partial^2K}{\partial V^2}\right)_{V=0} \left(\frac{\partial_\mu \phi}{2N_f} +A_\mu\right)^2 + \cdots \ .
\end{eqnarray}

The introduced D-term is
\begin{eqnarray}
V_D = \frac{g^2}{2}\left(\frac{-N_f}{4\pi^2} \partial_T K + \sum_i q_i \phi_i \partial_{\phi_i}K \right)^2 \ ,
\end{eqnarray}
where $\phi_i$ are all the fields that linearly couple to the $U(1)$ and $q_i$ are their charges.
To go further, we need a detailed form of the K\"ahler potential and the gauge kinetic function, whose string origin will be discussed in the next section, but even at this point it seems possible that $T + T^\dagger$ and hence the FI-term would be stabilized for a judicious choice of the K\"ahler potential.

As a side remark, we should point out that the stabilization of the non-zero FI-term and the D-term SUSY breaking is only possible with the help of the F-term SUSY breaking. This is due to the complexification of the gauge symmetry in the SUSY field theories which always enables the modulus of the charged field to be adjusted so that we have effectively zero D-term. In the supergravity theory, we also have a relation \cite{Joichi:1994tq}\cite{Choi:2005ge} 
\begin{eqnarray}
\sum_i\delta\phi_i \frac{D_i W}{W} = D \ ,
\end{eqnarray}
where $\delta\phi_i$ is a gauge transformation of the matter field.
As a consequence, unless $W=0$, it is impossible to obtain D-term SUSY breaking without F-term SUSY breaking.

\section{Moduli stabilization in stringy ISS model}
In section 2, we discussed the moduli stabilization of the ISS model from the field-theory viewpoints. To reach a definite conclusion, we need a detailed form of the K\"ahler potential, which depends on the UV physics. We propose a stringy realization of the ISS model, with a slight modification necessary for the moduli stabilization as discussed in the previous section, to examine the moduli stabilization with a metastable de-Sitter vacuum. 

Our stringy setup, which is inspired by \cite{Cremades:2007ig}, is as follows. We consider type IIB superstring with flux compactification (see e.g. \cite{Douglas:2006es} for a review). All the complex structure moduli and the dilaton are assumed to be fixed by the flux. For simplicity, we focus on the Calabi-Yau orientifold compactification \cite{Kachru:2003aw} with one K\"ahler modulus (which will be denoted by $T$), but generalizations to multi-moduli compactification should be possible. 

We consider a set of D7-branes and O-planes. We introduce a magnetic flux turned on in one of the D7-branes wrapped around the 4-cycle corresponding to the K\"ahler modulus $T$, which will generate a FI term in the effective four-dimensional field theory \cite{Jockers:2005zy}. The remaining D7-branes will give $SU(N_c)$ super Yang-Mills theory. The total gauge group is thus $U(1)\times SU(N_c)$. 

The matter contents of the low-energy effective field theory can be summarized as follows:
\begin{itemize}
	\item The field $\varphi$ stretching between the magnetized brane and $SU(N_c)$ branes will be charged $(+1,{N}_c)$ under $U(1)\times SU(N_c)$. 
	\item The field $\bar{\varphi}$ stretching between the magnetized brane and the orientifold images of $SU(N_c)$ branes will be charged $(+1,\bar{N}_c)$ under $U(1)\times SU(N_c)$. 
	\item The field $\rho$ stretching between the magnetized brane and its orientifold images will be charged $- 2$ under $U(1)$.
\end{itemize}
To obtain the ISS model precisely, we have to introduce multiple copies of magnetized D-branes to generate additional $SU(N_f)$ flavor symmetries. The moduli of $SU(N_f)$ flavor groups will be fixed at a sufficiently small coupling to be regarded as a global symmetry. The moduli fixing here has no theoretical difficulty but only makes the problem slightly complicated, so we simply assume that this is the case (see, however, some related discussions in \cite{Forste:2006zc}\cite{Giveon:2007fk}\cite{Amariti:2007am}).

So far, we have obtained all the matter ingredients to realize the stringy ISS model with the mass moduli $\rho$ and the K\"ahler moduli $T$. The string interaction gives a superpotential
\begin{eqnarray}
W = \rho \varphi_i \bar{\varphi}^i \ ,
\end{eqnarray}
and a necessary D-term interaction including the dynamical FI term that comes from the Chern-Simons coupling
\begin{eqnarray}
\int_{D7} C_4\wedge F\wedge F \ .
\end{eqnarray}
When an appropriate topological condition is satisfied, the modulus $T$ will be charged under the $U(1)$ and cancels the anomaly of the low-energy effective field theory \cite{Cremades:2007ig}.

For $N_c < N_f < \frac{3}{2}N_c$, the low energy physics is better studied in the magnetic dual description. As explained in section 2, we introduce dual quarks $q$ and $\bar{q}$ together with singlet mesons $M$. The effective four-dimensional supergravity has the superpotential\footnote{We have neglected the non-perturbative term here since we focus on the metastable ISS-type vacua along the same reasoning in the field theory analysis \cite{Intriligator:2006dd}. The non-perturbative term would be important to discuss the SUSY preserving vacua, and around such vacua, after integrating out the massive flavors, the moduli stabilization problem reduces to the one studied in \cite{Cremades:2007ig}.}
\begin{eqnarray}
W = W_0 + \rho \mathrm{Tr}M + \frac{1}{\mu}q^iM_{ij} \bar{q}^j
\end{eqnarray}
with the K\"ahler potential ($2\tau = T+T^\dagger$)
\begin{eqnarray}
K = -2\log(\tau^{3/2} + \zeta) + \frac{|\rho|^2}{\tau^{n}} + \frac{|q|^2 + |\bar{q}|^2}{\tau^{n}} + \frac{|M|^2}{\tau^{n}}e^{\frac{8\pi^2 \tau}{3N_c-2N_f}} \ . 
\end{eqnarray}
Here $W_0$ is the (constant) flux superpotential, and $\zeta$ is the stringy $\alpha'$ correction to the K\"ahler potential, which is proportional to the Euler number of the Calabi-Yau \cite{Becker:2002nn}. The modular weight $-n$ is not known in the magnetic description, so we will take it as a free parameter for a moment.\footnote{In the electric setup discussed in \cite{Cremades:2007ig}, they used $n=2/3$ from the result \cite{Conlon:2006tj} for chiral matters.}
 Finally, we have used the relation between the Landau pole scale $\tilde{\Lambda}$ and $T$: $\tilde{\Lambda}^{2N_f-3N_c} = e^{+8\pi^2 T} $.

The total potential is given by
\begin{eqnarray}
V = V_F + V_D \ ,
\end{eqnarray}
where the supergravity F-term potential gives
\begin{eqnarray}
V_F = e^{K} (K^{i\bar{j}}D_iW \bar{D}_{\bar{j}}\bar{W} - 3|W|^2) \ , \label{Fterm}
\end{eqnarray}
and the D-term potential gives
\begin{eqnarray}
V_D &=& \frac{1}{2\tau} \left(\frac{3N_f}{8\pi^2\tau}(1+\zeta\tau^{-3/2})^{-1} - \frac{2|\rho|^2+ |q|^2+|\bar{q}|^2 -2|M|^2 e^{\frac{8\pi^2\tau}{3N_c-2N_f}}}{\tau^{n}} \right. \cr & & + \left. \frac{N_fn(|\rho|^2+|q|^2+|\bar{q}|^2 + |M|^2 e^{\frac{8\pi^2\tau}{3N_c-2N_f}})}{4\pi^2 \tau^{n+1}} - \frac{2N_f|M|^2e^{\frac{8\pi^2\tau}{3N_c-2N_f}}}{(3N_c-2N_f)\tau^n}  
\right)^2 \ . \label{Dterm}
\end{eqnarray}
Note that we have used the fact that in our D7-brane model, the gauge kinetic term is dominated by $\tau$ as 
\begin{eqnarray}
\frac{2}{g^2} =  T + T^\dagger + (\mathrm{small \ dilaton \ contribution}) \ .
\end{eqnarray}

As can be inferred from the field-theory discussion in the last section, the central issue is how the K\"ahler moduli $\tau$ is stabilized. To see a general picture of the potential in the large $\tau$ regime, we first note that the leading order contribution comes from the D-term \eqref{Dterm} as $\frac{1}{2\tau}\left(\frac{3N_f}{8\pi^2\tau}\right)^2$, which shows a runaway behavior. The sub-leading contribution to the matter-independent potential for $\tau$ comes from the supergravity F-term \eqref{Fterm} as $\frac{3\zeta}{2 \tau^{9/2}}|W_0|^2$ due to the approximate no-scale-type K\"ahler potential. 

At this point, there are two options. When $\zeta >0$, the higher $1/\tau$ correction terms should give a negative contribution to the potential. In this case, if one further assumes that the potential, in the small $\tau$ region, should become (positively) large enough to avoid a singular behavior, it is generically expected that we achieve metastable vacua with a finite $\tau$. When $\zeta <0$, only the assumption that the the potential grows near $\tau^{3/2} \sim -\zeta$ suffices to obtain metastable vacua. We will demonstrate both possibilities in the following.

To confirm the above general expectation, we need to extremize the full supergravity potential. At a first glance, it would be a technically difficult problem to find exact extrema of the total potential. 
Here, in the following, we would like to argue that the ISS vacuum with the mass moduli $\rho$ fixed can be one of the extrema of the potential in the large $\tau$ limit.\footnote{At the same time, we also assume $e^{-\frac{8\pi^2 \tau}{3N_c-2N_f}}$ is not so small to satisfy the ISS condition $|\rho| \ll |\Lambda|$.}

We first integrate out the ISS matter fields $q$, $\bar{q}$ and $M$ to obtain the effective potential for $\tau$ and $\rho$. In the large $\tau$ regime, the matter-dependent F-term potential \eqref{Fterm} scales as $\tau^{n-3} \sum |\partial_i W|^2 $ with no contribution from $W_0$ due to the approximate no-scale-type K\"ahler potential. As a consequence, to integrate out the ISS matter fields, the F-term dominates over the D-term  for $n>\frac{1}{2}$ while the D-term dominates over the F-term for $n<\frac{1}{2}$. 

Let us concentrate on the former case ($n>\frac{1}{2}$). At the first order approximation, one can minimize the F-term potential, which yields an ISS vacuum. We note that at the ISS vacuum, the field dependent superpotential vanishes and the superpotential is given by the flux part alone: $W|_{\mathrm{ISS}}=W_0$. Then, neglecting the supergravity corrections, which turns out to appear as higher $1/\tau$ corrections, we obtain a potential for $\rho$ as in \eqref{topote}:
\begin{eqnarray}
V(\rho) \sim N_c\tau^{n-3}|\rho|^2 e^{-\frac{8\pi^2\tau}{3N_c-2N_f}} + \frac{1}{2\tau}\left(\frac{3N_f}{8\pi^2}\frac{1}{\tau} - \frac{2|\rho|^2+2(N_f-N_c)|\mu\rho|}{\tau^n}\right)^2 \ .
\end{eqnarray}

 At this order, as is the case with the global SUSY limit discussed in the previous section, $\rho$ is fixed at
\begin{eqnarray}
|\rho| = \frac{\frac{3N_f(N_f-N_c)|\mu|}{8\pi^2 \tau^{n+2}}}{N_c\tau^{n-3} e^{ -\frac{8\pi^2 \tau}{3N_c-2N_f}} + \frac{2|\mu|^2(N_f-N_c)^2}{\tau^{2n+1}}} \ . \label{valrho}
\end{eqnarray}

To see the moduli stabilization of $\tau$, we note that in addition to the contribution from the potential in the global SUSY limit
\begin{eqnarray}
V_{\mathrm{global}} = \frac{1}{2\tau}\left(\frac{3N_f}{8\pi^2\tau}\right)^2 - \frac{(N_f-N_c)^2\left(\frac{3N_f}{8\pi^2\tau}\right)^2\frac{|\mu|^2}{\tau^{2n+2}}}{ N_c\tau^{n-3}e^{ -\frac{8\pi^2 \tau}{3N_c-2N_f}} + \frac{2|\mu|^2(N_f-N_c)^2}{\tau^{2n+1}}}  \ , 
\end{eqnarray}
we have supergravity F-term corrections
\begin{eqnarray}
\delta V_{\mathrm{SUGRA}} &=& \frac{3\zeta}{2 \tau^{9/2}}|W_0|^2 +(1-n-n^2)\frac{|\rho|^2 + 2(N_f-N_c)|\rho\mu|}{\tau^{3+n}}|W_0|^2 + \cdots \cr
&=& \frac{3\zeta}{2 \tau^{9/2}}|W_0|^2 + (1-n-n^2)\frac{2(N_f-N_c)^2\left(\frac{3N_f}{8\pi^2}\right)\frac{|\mu|^2|W_0|^2}{\tau^{2n+5}}}{N_c\tau^{n-3} e^{ -\frac{8\pi^2 \tau}{3N_c-2N_f}} + \frac{2|\mu|^2(N_f-N_c)^2}{\tau^{2n+1}}} + \cdots \ . 
\end{eqnarray}
The D-term corrections and higher K\"ahler corrections are neglected here. Substituting \eqref{valrho}, we obtain a total potential $V_{\mathrm{global}} + \delta V_{\mathrm{SUGRA}}$ for $\tau$. It is possible to stabilize $\tau$ in a metastable de-Sitter vacuum with a vanishingly small cosmological constant by an appropriate fine-tuning of flux parameters.

For example, by taking $N_c = 5000$, $N_f=6000$, $|\mu| = 14.72$, $\zeta = 143$, and  $|W_0| = 0.11$ for $n =1$, we will obtain a metastable vacuum as expected from the general argument for $\zeta > 0$. The moduli $\tau$ is fixed around $\tau \sim 1.4$ and the cosmological constant can be arbitrarily small. See figure 1 for the effective potential for $\tau$. For comparison, we have also shown the $(\rho,\tau)$ potential in figure 2. We can see the consistency of our approach because the mass for the $\rho$ field is much larger than that for $\tau$.

While we need a significant fine-tuning to obtain de-Sitter vacua, it is slightly easier to obtain anti de-Sitter vacua with relatively small number of branes.
 For instance, $N_c=10$, $N_f = 12$, $\zeta = 0.9$, $\mu=10^{-4}$, $|W_0|=0.01$ 
 for $n=1$ gives the anti de-Sitter vacuum $(V_0 \sim -2\times 10^{-10})$ with $\tau \sim 11$. If one assumes that the ISS vacuum continues to be a first order solution to the full potential for $n\le 1/2$, we will also obtain a metastable vacuum for $\zeta <0$ in our approximation by taking e.g. $N_c = 10$, $N_f=12$, $|\mu| = 0.00008$, $\zeta = -6$, and  $|W_0| = 0.01$ for $n =1/2$ with $\tau \sim 2.2$ and $V_0 = -2 \times 10^{-5}$. In these cases, we need a suitable uplifting mechanism to obtain a de-Sitter vacuum.

\begin{figure}[htbp]
    \begin{center}
    \includegraphics[width=0.6\linewidth,keepaspectratio,clip]
      {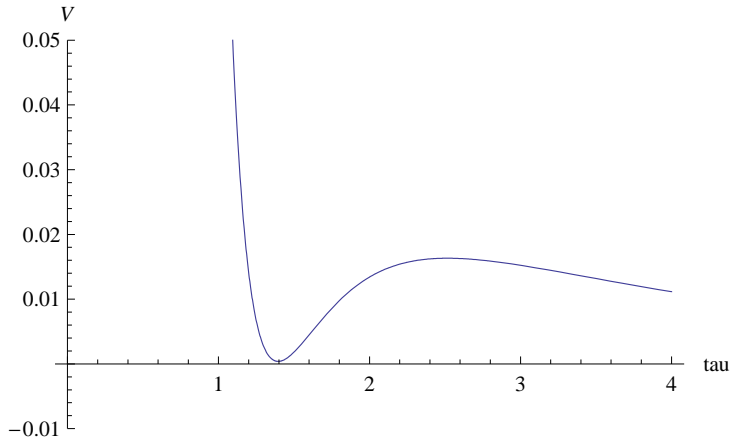}
    \end{center}
    \caption{Effective potential for $\tau$. The cosmological constant can be tuned to be arbitrarily small.}
    \label{fig1}
\end{figure}

\begin{figure}[htbp]
    \begin{center}
    \includegraphics[width=0.6\linewidth,keepaspectratio,clip]
      {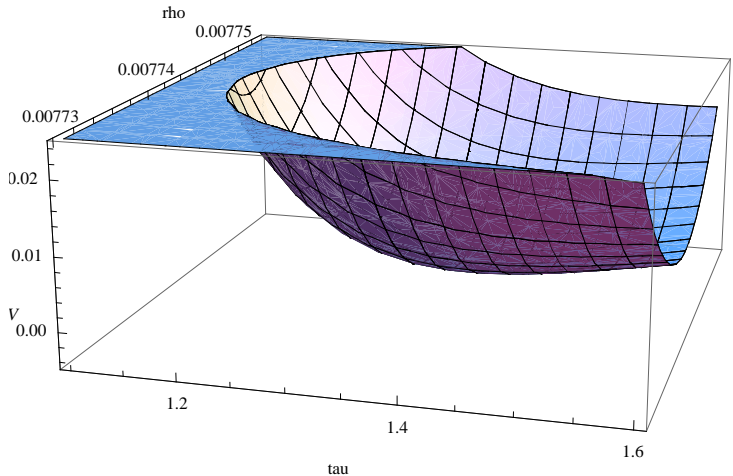}
    \end{center}
    \caption{Potential for $(\rho,\tau)$. The mass moduli $\rho$ is much heavier than $\tau$, so our approximation is consistent. The minimum of the potential is located at almost the same point as in figure 1.}
    \label{fig2}
\end{figure}

\section{Discussion}
In this paper, we have investigated the moduli stabilization problem of the dynamical (metastable) SUSY breaking models in the string flux compactification. As a concrete example, we have used the ISS(-like) model with an extra $U(1)$ gauge symmetry.
The moduli stabilization is well under control for large $\tau$ (but not too large $\tau$). 

In our construction, the SUSY breaking scale is naturally related with the string scale unless we allow a rather significant amount of  fine-tuning of parameters. One possibility to obtain a low energy SUSY breaking more naturally is to use a warped compactification, which would drastically reduce the relevant energy scale of the physics on the 4-cycle where the SUSY breaking occurs. 

As in the original ISS model, our model has several Goldstone modes, part of which are absorbed into massive vector multiplets due to the Higgs mechanism. In addition, the mass spectrum is hierarchical and we have a light moduli field $\tau$, typically with the gravitino mass scale, when the SUSY breaking is sufficiently small. The cosmological significance and constraint on these light fields would be of most importance from phenomenological viewpoints.

Among other things, one big feature of our model is that it naturally accompanies the D-term SUSY breaking as well as the F-term SUSY breaking due to the dynamical FI term. Recently, a phenomenologically interesting class of D-term gauge mediation has been proposed \cite{Nakayama:2007cf} with very light gravitino ($\sim O(1)$ eV) and a possible candidate for the dark matter \cite{Hamaguchi:2007rb}.
The stringy realization in \cite{Nakayama:2007cf} was in the global SUSY limit without giving an explicit moduli stabilization mechanism. Since our model fixes the dynamical FI term within the supergravity, which is known to be difficult in general (see e.g. \cite{Binetruy:2004hh}), applications to the strongly coupled D-term gauge mediation would be promising. For this, we need a better control of the potential in the small $\tau$ region. We expect that by using string dualities, the analysis for the small $\tau$ region would be possible.

\section*{Acknowledgements}
The research of Y.~N. is supported in part by NSF grant PHY-0555662 and the UC Berkeley Center for Theoretical Physics.

\end{document}